\PassOptionsToPackage{unicode}{hyperref}
\PassOptionsToPackage{hyphens}{url}
\PassOptionsToPackage{dvipsnames,svgnames,x11names}{xcolor}
\documentclass[
  12pt]{article}

\usepackage{amsmath,amssymb}
\usepackage{amsthm}
\usepackage{algorithm}
\usepackage{algorithmic}
\usepackage{iftex}
\theoremstyle{remark}
\newtheorem{remark}{Remark}

\ifPDFTeX
  \usepackage[T1]{fontenc}
  \usepackage[utf8]{inputenc}
  \usepackage{textcomp} 
\else 
  \usepackage{unicode-math}
  \defaultfontfeatures{Scale=MatchLowercase}
  \defaultfontfeatures[\rmfamily]{Ligatures=TeX,Scale=1}
\fi
\usepackage{lmodern}
\ifPDFTeX\else  
\fi
\IfFileExists{upquote.sty}{\usepackage{upquote}}{}
\IfFileExists{microtype.sty}{
  \usepackage[]{microtype}
  \UseMicrotypeSet[protrusion]{basicmath} 
}{}
\makeatletter
\@ifundefined{KOMAClassName}{
  \IfFileExists{parskip.sty}{%
    \usepackage{parskip}
  }{
    \setlength{\parindent}{0pt}
    \setlength{\parskip}{6pt plus 2pt minus 1pt}}
}{
  \KOMAoptions{parskip=half}}
\makeatother
\usepackage{xcolor}
\makeatletter
\ifx\paragraph\undefined\else
  \let\oldparagraph\paragraph
  \renewcommand{\paragraph}{
    \@ifstar
      \xxxParagraphStar
      \xxxParagraphNoStar
  }
  \newcommand{\xxxParagraphStar}[1]{\oldparagraph*{#1}\mbox{}}
  \newcommand{\xxxParagraphNoStar}[1]{\oldparagraph{#1}\mbox{}}
\fi
\ifx\subparagraph\undefined\else
  \let\oldsubparagraph\subparagraph
  \renewcommand{\subparagraph}{
    \@ifstar
      \xxxSubParagraphStar
      \xxxSubParagraphNoStar
  }
  \newcommand{\xxxSubParagraphStar}[1]{\oldsubparagraph*{#1}\mbox{}}
  \newcommand{\xxxSubParagraphNoStar}[1]{\oldsubparagraph{#1}\mbox{}}
\fi
\makeatother

\usepackage{longtable,booktabs,array}
\usepackage{calc} 
\usepackage{etoolbox}
\makeatletter
\patchcmd\longtable{\par}{\if@noskipsec\mbox{}\fi\par}{}{}
\makeatother
\IfFileExists{footnotehyper.sty}{\usepackage{footnotehyper}}{\usepackage{footnote}}
\makesavenoteenv{longtable}
\usepackage{graphicx}
\makeatletter
\def\maxwidth{\ifdim\Gin@nat@width>\linewidth\linewidth\else\Gin@nat@width\fi}
\def\maxheight{\ifdim\Gin@nat@height>\textheight\textheight\else\Gin@nat@height\fi}
\makeatother
\setkeys{Gin}{width=\maxwidth,height=\maxheight,keepaspectratio}
\makeatletter
\def\fps@figure{htbp}
\makeatother

\makeatletter
\@ifpackageloaded{caption}{}{\usepackage{caption}}
\AtBeginDocument{%
\ifdefined\contentsname
  \renewcommand*\contentsname{Table of contents}
\else
  \newcommand\contentsname{Table of contents}
\fi
\ifdefined\listfigurename
  \renewcommand*\listfigurename{List of Figures}
\else
  \newcommand\listfigurename{List of Figures}
\fi
\ifdefined\listtablename
  \renewcommand*\listtablename{List of Tables}
\else
  \newcommand\listtablename{List of Tables}
\fi
\ifdefined\figurename
  \renewcommand*\figurename{Figure}
\else
  \newcommand\figurename{Figure}
\fi
\ifdefined\tablename
  \renewcommand*\tablename{Table}
\else
  \newcommand\tablename{Table}
\fi
}
\@ifpackageloaded{float}{}{\usepackage{float}}
\floatstyle{ruled}
\@ifundefined{c@chapter}{\newfloat{codelisting}{h}{lop}}{\newfloat{codelisting}{h}{lop}[chapter]}
\floatname{codelisting}{Listing}

\makeatother
\makeatletter
\@ifpackageloaded{caption}{}{\usepackage{caption}}
\@ifpackageloaded{subcaption}{}{\usepackage{subcaption}}
\makeatother

\ifLuaTeX
  \usepackage{selnolig}  
\fi
\usepackage[]{natbib}
\usepackage{bookmark}

\usepackage{tikz}
\usetikzlibrary{shapes.geometric} 

\IfFileExists{xurl.sty}{\usepackage{xurl}}{} 
\hypersetup{
  pdftitle={Title},
  pdfauthor={Author 1; Author 2},
  pdfkeywords={3 to 6 keywords, that do not appear in the title},
  colorlinks=true,
  linkcolor={blue},
  filecolor={Maroon},
  citecolor={Blue},
  urlcolor={Blue},
  pdfcreator={LaTeX via pandoc}}

\newcommand{\anon}{1}



\begin{document}

\def\spacingset#1{\renewcommand{\baselinestretch}%
{#1}\small\normalsize} \spacingset{1}


\if1\anon
{
  \title{\bf The Bag-and-Whisker Plot: A New Bagplot for Bivariate Data}
  \author{
    Shenghao Qin\textsuperscript{1} \and
    Bowen Gang\textsuperscript{1} \and
    Tiejun Tong\textsuperscript{2} \and
    Hengjian Cui\textsuperscript{3}
  }
  \date{%
    \small
    \textsuperscript{1}Department of Statistics and Data Science, Fudan University\\
    \textsuperscript{2}Department of Mathematics, Hong Kong Baptist University\\
    \textsuperscript{3}School of Mathematical Sciences, Capital Normal University
  }
  \maketitle
} \fi

\if0\anon
{
  \bigskip
  \bigskip
  \bigskip
  \begin{center}
    {\LARGE\bf The Bag-and-Whisker Plot: A New Bagplot for Bivariate Data}
\end{center}
  \medskip
} \fi

\bigskip
\begin{abstract}
The bagplot, also known as the ``bag-and-bolster plot'', is a notable extension of the boxplot from univariate to bivariate data. Although widely used, its practical application is hindered by two key limitations: the fixed inflation factor for outlier detection that does not adapt to the sample size, and the unstable convex hull used to visualize its fence.
 In this paper, we propose a new bagplot, namely the ``bag-and-whisker plot'', as an improvement method to address these limitations. Our framework recasts outlier detection as a multiple testing problem, yielding a data-adaptive fence that controls statistical error rates  and enhances the reliability of outlier identification. To further resolve graphical instability, we introduce a refined visualization that abandons the convex hull (the bolster) with a direct rendering of the statistical fence, complemented by granular whiskers that effectively illustrate the data's spread. 
Extensive simulations and real-world data analyses demonstrate that our new bagplot exhibits superior adaptivity and robustness compared to the existing standard, and thus can be highly recommended for practical use. To increase the visibility of the work, a user-friendly R package named \texttt{BagWhiskerPlot} has been made publicly available on CRAN.
 


\end{abstract}

\noindent%
\vfill

\newpage
\spacingset{1.8} 

\section{Introduction}\label{sec-intro}

Tukey's box-and-whisker plot is a foundational tool in exploratory data analysis, celebrated for its robust and concise summary of a dataset's key features \citep{tukey1977exploratory}. A defining element is its rule for flagging potential outliers, which designates any point beyond 1.5 times the interquartile range (IQR) from the central box as suspect. This 1.5$\times$IQR heuristic, while simple and effective for its time, was not derived from formal statistical principles and its performance is known to degrade in large samples, often flagging an excessive number of points. For decades, numerous modifications to the inflation factor of 1.5 were proposed, representing significant efforts to refine the heuristic approach for outlier detection  \citep{hoaglin1986performance,hoaglin1987fine,sim2005outlier}.
Recent advancements have addressed this by introducing a new generation of boxplots with data-adaptive fences. These newer methods, such as the Chauvenet-type, Holm-type, and BH-type boxplots, adjust their outlier detection thresholds based on the sample size, offering a more statistically principled and reliable framework for univariate data analysis \citep{gang2025box,lin2025tukey,tong2026chauboxplot}.


Just as the univariate boxplot evolved, so too did efforts to generalize it to bivariate data. 
This task, however, is fundamentally more complex due to the lack of a natural ordering of points in two dimensions \citep{wickham201140}. Early attempts, such as the rangefinder plot \citep{becketti1987rangefinder}, were simple constructions based on marginal statistics, effectively treating the variables as independent. 
The 1990s witnessed a proliferation of more sophisticated proposals, ranging from model-based elliptical plots \citep{goldberg1992bivariate} to non-parametric approaches based on convex hull peeling \citep{zani1998robust}.

The key conceptual advance that led to a widely adopted solution was the use of data depth. The bagplot, introduced by \citet{rousseeuw1999bagplot}, was built specifically on the concept of halfspace depth \citep{tukey1975mathematics}, while the broader utility of depth functions for robust multivariate graphics and inference was concurrently formalized in the influential work of \citet{liu1999multivariate}. The bagplot's construction is analogous to the univariate boxplot: a central ``bag'' contains the 50\% of data points with the maximum depth, and an outer ``fence'' is used for outlier detection. 

The bagplot has since established itself as the \emph{de facto} standard for 2D data visualization, a status attested to by over 850 citations in Google Scholar as of May 2026 and its permeation into diverse applied disciplines.
This widespread adoption underscores a critical and pervasive demand for tools that can intuitively reveal bivariate relationships, non-linear associations, and outlier patterns. In ecological and environmental science, for instance, \cite{schirpke2019integrating} employed bagplots to unravel trade-offs and synergies among ecosystem services, leveraging the tool to visualize data dispersion that linear methods overlook. In physical anthropology, \cite{o2012diet} utilized the method to clarify isotopic offsets in paleodietary studies, relying on the visualization of data concentration to make accurate interpretations. The tool's cross-disciplinary relevance is further exemplified in environmental epidemiology, where \cite{damialis2021higher} used bagplots to confirm correlations between airborne pollen and SARS-CoV-2 infection rates across 31 countries, and in neuroscience, where \cite{baliki2011brain} applied them to classify brain morphological signatures in chronic pain.

However, the contrast between this extensive cross-disciplinary demand and the bagplot's stagnant methodological development reveals a significant opportunity. While the bagplot is the prevailing standard, its original formulation relies on heuristics that have not kept pace with modern statistical rigor. Specifically, it suffers from two critical limitations. First, its outlier detection rule uses a fixed inflation factor (typically 3) that fails to adapt to the  sample size or dimension—a static heuristic similar to the univariate $1.5 \times \text{IQR}$ rule that modern methods have since superseded. Second, its fence visualization, known as the ``loop'', is defined as the convex hull of non-outlying points.
\cite{rousseeuw1999bagplot} noted that this loop effectively plays the role of a bolster, suggesting that their method could be more accurately termed a ``bag-and-bolster plot'' rather than a box-and-whisker equivalent. 
 This construction is empirically unstable and often visually misleading, as the hull's shape is highly sensitive to the specific locations of peripheral points, creating a disconnect between the statistical rule and its graphical representation.

In this paper, we enhance the bagplot by addressing these statistical and visual limitations, effectively transforming the ``bag-and-bolster plot'' into a true ``bag-and-whisker plot''. 
We first incorporate the principles of multiple testing, replacing the fixed inflation factor with a data-adaptive method. This allows for the construction of bagplots with fences that are inspired by controlling various error rates. To resolve the visual instability of the bolster (loop), we propose a new graphical approach. Our method visualizes a theoretical fence, which is a stable contour based on the chosen statistical threshold, alongside a new implementation of granular whiskers. This dual representation provides a more robust interpretation by clearly distinguishing the underlying statistical boundary from the empirical shape of the data, creating a more reliable and informative tool for bivariate exploration.


\section{The Bagplot by \cite{rousseeuw1999bagplot}}
\label{sec:review_bagplot}

A bivariate generalization of the univariate boxplot, the bagplot, developed by \citet{rousseeuw1999bagplot}, serves as a tool for visualizing essential characteristics of a dataset, including its location, spread, correlation, and skewness. Its construction is founded on the concept of halfspace depth \citep{tukey1975mathematics}, a multivariate extension of the univariate notion of rank.

\subsection{Construction of the Original Bagplot}
\label{subsec:construction}
For a given bivariate dataset $\mathcal{Z} = \{\boldsymbol{z}_1, \dots, \boldsymbol{z}_n\}$, where $\boldsymbol{z}_i \in \mathbb{R}^2$, the halfspace depth of a point $\boldsymbol{\theta} \in \mathbb{R}^2$ is the minimum number of data points contained in any closed half-plane with a boundary passing through $\boldsymbol{\theta}$. The bagplot is constructed through the following steps:

\begin{enumerate}
	\item \textbf{The Depth Median:} The point with the maximum halfspace depth is identified as the depth median, denoted $\boldsymbol{\mu}_{\text{depth}}$. This serves as the bivariate analogue of the univariate median, providing a robust estimate of the data's center.
	
	\item \textbf{The Bag:} A central polygon, called the ``bag'', is constructed to contain about 50\% of observations with the maximum depth. The bag is the two-dimensional counterpart to the IQR and robustly captures the data's spread and shape.
	
	\item \textbf{The Fence:} A theoretical outer boundary, called the ``fence'', is determined by magnifying the bag by a factor of 3 relative to the depth median. Any point lying outside this fence is flagged as a potential outlier.
	
	\item \textbf{The Loop:} The region between the bag and the fence contains the remaining non-outlying points. The bagplot visualization does not typically render the fence directly. Instead, the boundary of the non-outlier region is depicted by the convex hull of all points contained within the loop (including the bag itself).
\end{enumerate}

\subsection{Limitations of the Original Bagplot}
\label{subsec:limitations}
While the bagplot is a powerful exploratory tool, its original formulation presents two key limitations that can affect its statistical and visual interpretation.

\subsubsection{Fixed Inflation Factor}
The use of 3 as the inflation factor for the bagplot is a fixed heuristic. This can be contrasted with the univariate boxplot, where the 1.5$\times$IQR rule for the two fences is a well-established convention. A notable property of this univariate rule is that, for data drawn from a normal distribution, it flags approximately 0.7\% of observations as outliers. Under a bivariate normal distribution, however, the bagplot's inflation factor of 3 results in only $0.195\%$ of observations being outside of the fence. This proportion is substantially more conservative than the 0.7\% benchmark, highlighting a clear inconsistency between the methods.

The challenge of selecting an appropriate inflation factor has been previously noted in the literature. For instance, \citet{zani1998robust}, in their work on bivariate boxplots based on convex hull peeling and B-spline smoothing, derived an inflation factor of 1.58 specifically to achieve an outlier percentage of approximately 1\% under bivariate normality. While this represents a principled attempt to standardize the outlier rate, the resulting factor is still a fixed constant. As \citet{zani1998robust} further observed, for any fixed factor, the proportion of flagged points from a normal sample is a decreasing function of the sample size $n$. This underscores the fundamental limitation of any fixed-factor approach.
Furthermore, this discrepancy is compounded by dimensionality. The need for an inflation factor to decrease as the data's dimension increases, in order to maintain a consistent outlier proportion of 0.7\%, demonstrates the inadequacy of a single fixed value. The inflation factors required to achieve this 0.7\% proportion, up to 10 dimensions, are shown in Table~\ref{tab:inflation_factors}.
\begin{table}[h!]
	\centering
	\caption{Inflation factors required to maintain a constant outlier proportion of 0.7\% (the benchmark from a standard univariate boxplot) as dimensionality grows.}
	\label{tab:inflation_factors}
	\begin{tabular}{@{}lcccccccccc@{}}
		\toprule
		Dimension         & 1    & 2    & 3    & 4    & 5 &6 &7&8&9&10   \\ \midrule
		Inflation Factor  & 4.00 & 2.68 & 2.26 & 2.05 & 1.91&1.82&1.75&1.69&1.65&1.61 \\ \bottomrule
	\end{tabular}
\end{table}

\subsubsection{Fence Visualization}\label{limitation:loop}
The second limitation of the bagplot arises from a mismatch between its statistical definition and its graphical implementation. The outlier detection rule is based on a theoretical fence, which a simple magnification of the central bag. However, the visualization does not show this theoretical boundary. Instead, it displays the convex hull of all points not flagged as outliers. This practice creates a visually unstable and potentially misleading representation for two reasons.

\begin{figure}[h!]
	\centering
		\includegraphics[width=0.9\textwidth]{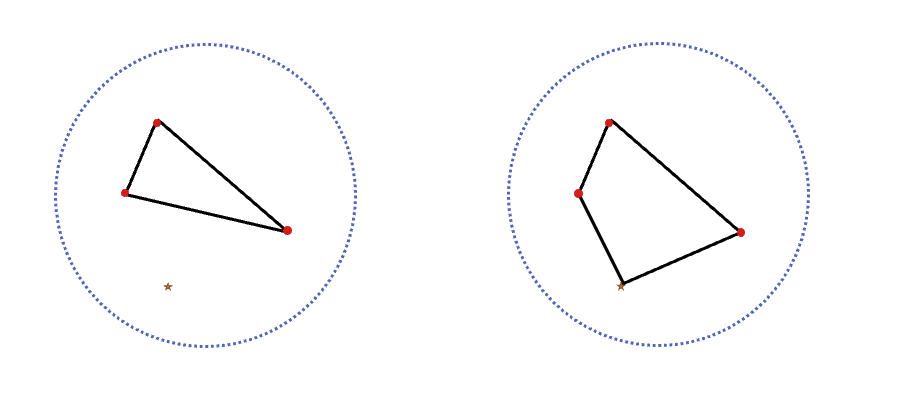}
	\caption{Visual limitations of the ``bolster'' (convex hull) representation in the classic bag-and-bolster plot by \cite{rousseeuw1999bagplot}. Left: The star ($\star$) lies strictly inside the theoretical fence (dashed contour), yet it falls outside the visualized convex hull (solid polygon), leading to incorrect outlier inference. Right: The shape of the convex changes drastically with the inclusion of a single point at the star's position, thus altering the entire shape of the visualized boundary.}
	\label{fig:misleading_hull}
\end{figure}

First, it creates a visual paradox not present in the univariate boxplot. In a standard boxplot, the fence is simply the end of the whisker. Any data point lying between the median and the whisker is unambiguously inside the fence. In the bagplot, a point can be well within the theoretical fence but lie outside the visualized convex hull, as illustrated in the left panel of Figure~\ref{fig:misleading_hull}. The dashed line represents the theoretical fence boundary obtained by magnifying the bag. The solid polygon is the convex hull of the loop points, which is what is typically visualized. A point at the location of the star ($\star$) is inside the theoretical fence but outside the visualized convex hull.
An observer would incorrectly infer that a point at the star's location would be flagged as an outlier. In reality, the point's data depth may be no smaller than that of the vertices defining the convex hull; according to the statistical rule, it is not an outlier.

Second, this visualization is empirically unstable. The shape of the convex hull is highly sensitive to the exact location of its defining vertices. The addition, removal, or slight perturbation of a single data point near the periphery can dramatically and non-locally alter the shape of the visualized boundary. As illustrated in the right panel of Figure~\ref{fig:misleading_hull}, if there is a data point at the star's position, the shape of the convex hull will change drastically.
This instability means the graph is not a robust representation of the data's underlying distribution.
This disconnect between the formal rule and the visual evidence undermines the plot's reliability. The graphic should be a faithful representation of the statistical procedure, a principle that the conventional bagplot violates.


\section{New Bagplot: The Bag-and-Whisker Plot}
\label{sec:methodology}

In this section, we propose a comprehensive enhancement to the bagplot that addresses the limitations of both its fixed inflation factor and its fence visualization. Our methodology consists of two primary contributions. First, we introduce a formal multiple testing framework to construct a data-adaptive fence. Second, 
we redesign the visualization of spread and skewness by  replacing the unstable ``bolster'' with explicit whiskers, effectively transforming the original ``bag-and-bolster'' display into a genuine bivariate bag-and-whisker plot.

\subsection{A Multiple Testing Framework for Adaptive Inflation Factor}
\label{subsec:framework}
We recast the bagplot's outlier detection rule as a formal multiple testing procedure, analogous to recent work unifying the univariate boxplot \citep{gang2025box}. We treat the classification of each data point as a distinct hypothesis test. For each point $\boldsymbol{z}_i$ in a dataset $\mathcal{Z} \subset \mathbb{R}^2$, the null hypothesis, $H_{0i}$, states that the point is drawn from the same underlying distribution as the bulk of the data. Flagging $\boldsymbol{z}_i$ as an outlier is thus equivalent to rejecting $H_{0i}$.

The original bagplot, with its inflation factor of 3, corresponds to an unadjusted testing procedure that rejects $H_{0i}$ if its associated $p$-value (under the null hypothesis that the majority of the data follow a bivariate normal distribution) is less than approximately $0.00195$. 
This fixed threshold presents a dual limitation. In small samples, this cutoff is substantially more conservative than the $0.007$ benchmark of the standard univariate boxplot, resulting in overly strict control that may fail to detect genuine outliers. Conversely, in large samples, the lack of adjustment for multiplicity causes the expected number of false positives to grow linearly with the sample size, potentially resulting in an excess of flagged outliers.
Our solution is to replace this fixed-cutoff rule with a general and adaptive $p$-value-based pipeline. This approach allows for the application of standard multiple testing procedures to control error rates such as the Per-Family Error Rate (PFER), the Family-Wise Error Rate (FWER), or the False Discovery Rate (FDR) by \cite{benjamini1995controlling}, resulting in a bagplot with a statistically principled fence.

Our methodology is built on the pragmatic assumption that the central body of the data follows a bivariate normal distribution with mean $\mu$ and covariance matrix $\Sigma$. This approach is rooted in Winsor's principle that ``all distributions are normal in the middle'' \citep{tukey1960survey}, providing a robust foundation for identifying points that deviate from this central tendency. Our pipeline proceeds in the following steps:


\textbf{1. Robust Parameter Estimation.}
The first step of our pipeline requires robust estimation of the parameters $(\boldsymbol{\mu}, \boldsymbol{\Sigma})$ that characterize the central body of the data under a bivariate normal assumption. To ensure these estimates are resistant to observations distant from the main data cloud, we employ distinct methods for the location and scatter parameters. Specifically, we estimate the location vector $\boldsymbol{\mu}$ with the depth median. In cases where the point achieving the maximum halfspace depth is not unique, we define the estimator $\hat{\boldsymbol{\mu}}$ as the coordinate-wise median of the set of deepest points. For the covariance matrix $\boldsymbol{\Sigma}$, we employ the highly robust Minimum Covariance Determinant (MCD) estimator \citep{rousseeuw1984least}. The resulting estimates are denoted by $\hat{\boldsymbol{\mu}}$ and $\hat{\boldsymbol{\Sigma}}$, respectively.

\textbf{2. Bag Construction.}
Following the standard definition in \cite{rousseeuw1999bagplot}, we construct the central ``bag'' to contain about 50\% of observations with the maximum halfspace depth.

\textbf{3. $p$-value Calculation.}
To obtain a $p$-value for each point $z_i$, we first measure its standardized distance from the robustly estimated center using the squared Mahalanobis distance:
\begin{equation}
	d_i^2 = (\boldsymbol{z}_i - \hat{\boldsymbol{\mu}})^\top \hat{\boldsymbol{\Sigma}}^{-1}(\boldsymbol{z}_i - \hat{\boldsymbol{\mu}}).
	\label{eq:mahalanobis}
\end{equation}
Under the null hypothesis $H_{0i}$ that an observation $\boldsymbol{z}_i$ is not an outlier, its squared Mahalanobis distance can be compared to a $\chi^2_2$ distribution. This reference is based on the standard result that, assuming the bulk of the data follows a bivariate normal distribution, the squared robust Mahalanobis distance converges asymptotically to a $\chi^2_2$ distribution as $n \to \infty$ \citep{hardin2005distribution}. 
 The resulting $p$-value is the upper-tail probability:
\begin{equation}
	p_i = P(\chi_2^2 \ge d_i^2).
	\label{eq:pvalue}
\end{equation}

\textbf{4. Multiple Testing Adjustment.}
With the set of $p$-values $\{p_1, \dots, p_n\}$, we apply a standard multiple testing procedure to control a desired error rate (e.g., FWER, FDR, PFER) at a specified level $q$. This yields a data-dependent significance threshold, $t_{\text{adj}}$. Any hypothesis $H_{0i}$ with $p_i \le t_{\text{adj}}$ will be rejected. We then convert this threshold back to the critical value for the squared Mahalanobis distance:
\begin{equation}
	d^2_{\text{adj}} = F^{-1}_{\chi^2_2}(1 - t_{\text{adj}}),
	\label{eq:d_adj}
\end{equation}
where $F^{-1}_{\chi^2_2}$ is the quantile function of the $\chi^2_2$ distribution.

\textbf{5. Data-Adaptive Inflation Factor.}
Finally, we derive a data-adaptive inflation factor, $\lambda$, that ensures the visualized fence is both statistically principled and geometrically consistent with the multiple testing results. This is achieved through a two-stage process.

First, we define a \emph{statistical} inflation factor, $\lambda_{\text{stat}}$, that aligns the fence with the statistical threshold $d_{\text{adj}}^2$ from Equation~(3). Since the bag's boundary represents the 50th percentile of data depth, it should correspond approximately to the median of the squared Mahalanobis distances. We thus define $\lambda_{\text{stat}}$ as:
\begin{equation*}
	\lambda_{\text{stat}} = \sqrt{\frac{d_{\text{adj}}^2}{\text{median}(d_1^2, \dots, d_n^2)}}.
\end{equation*}
This factor scales the bag such that its boundary would, under ideal elliptical symmetry, align with the multiple testing cutoff.

However, the bag is a data-dependent polygon, not a perfect ellipse corresponding to an isocontour of the Mahalanobis distance. Consequently, magnifying the bag by $\lambda_{\text{stat}}$ does not guarantee that all non-rejected points will lie inside the resulting fence. To resolve this potential contradiction between the statistical test and the visual representation, we introduce a second, \emph{data-driven} inflation factor, $\lambda_{\text{data}}$, which is defined as the minimum magnification required to geometrically enclose all points $\boldsymbol{z}_i$ for which the null hypothesis $H_{0i}$ was not rejected. More precisely, let $\boldsymbol{b}_i$ be the intersection of the bag's boundary with the ray originating from $\hat{\boldsymbol{\mu}}$ and passing through $\boldsymbol{z}_i$. We define $\lambda_{\text{data}}$ as the maximum ratio of distances for all non-rejected points:
\begin{equation*}
	\lambda_{\text{data}} = \max_{i \in \mathcal{S}} \frac{\| \boldsymbol{z}_i - \hat{\boldsymbol{\mu}} \|}{\| \boldsymbol{b}_i - \hat{\boldsymbol{\mu}} \|},
\end{equation*}
where $\mathcal{S}$ is the set of indices for points where $H_{0i}$ is not rejected. 
The final data-adaptive inflation factor $\lambda$ is chosen to satisfy both the statistical criterion and the geometric constraint:
\begin{equation}\label{lambda}
	\lambda = \max(\lambda_{\text{stat}}, \lambda_{\text{data}}).
\end{equation}
This factor is then used to construct the outer fence by magnifying the bag.  A few remarks are in order.

\begin{remark}
A key strength of the proposed framework is its modularity. While our implementation is instantiated with the halfspace depth median and the MCD estimator, these are not essential components. The pipeline is compatible with any robust estimators for the location and scatter parameters. Alternative notions of data depth (e.g., \citealp{liu1999multivariate}) could be readily substituted for the location estimate. The MCD estimator is one of many powerful tools for robust covariance estimation, other possibilities include using the depth weighted scatter estimators by \cite{zuo2005depth} or the T-type estimator by \cite{tian2024abnormal}. For a comprehensive treatment of robust statistics, readers may refer to the  foundational texts such as \citet{hampel1986robust}, \citet{HuberRonchetti2009}, and \citet{maronna2019robust}.
\end{remark}

\begin{remark}\label{rmk2}
	We note that for finite samples, the distribution of squared robust Mahalanobis distances from an MCD estimator can be appropriately approximated by a scaled $F$-distribution when the data are from a normal distribution as shown by \citet{hardin2005distribution}. 
		 However, this approximation requires Monte Carlo calibration of the scaling parameters, which introduces substantial computational overhead. Given the computational cost we adopt the asymptotic $\chi^2$ reference as the default. For illustrative purposes, Sections~\ref{sec:examples} and \ref{sec:real-data} include bagplots constructed under both $p$-value constructions. Our accompanying R package \texttt{BagWhiskerPlot} implements both distributions for $p$-value constructions, allowing users to select the scaled $F$-distribution if they prefer.
		
\end{remark}

\begin{remark}
The necessity of our two-stage approach for deriving the inflation factor, $\lambda = \max(\lambda_{\text{stat}}, \lambda_{\text{data}})$, is a feature unique to the multivariate setting. In the univariate setting \citep{gang2025box}, a single-stage factor is sufficient. This is because a fundamental consistency exists in one dimension: the robust parameters estimates, the inner box and the fences are all derived directly from the same set of sample quantiles. The statistical rule and the graphical construction are perfectly aligned.
This direct alignment breaks down in two dimensions. Our statistical rule for outlier detection is based on the squared Mahalanobis distance, which defines perfectly elliptical contours of probability. The graphical construct, however, is the bag, which is a data-dependent convex polygon. There is no guarantee that this polygon is a perfect ellipse. Consequently, a simple magnification of the polygonal bag (the statistical target set by $\lambda_{\text{stat}}$) may not produce a fence that is geometrically consistent with the elliptical boundary defined by the multiple testing procedure. A point with a small Mahalanobis distance could lie in a direction where the bag has a ``flat side'', causing it to fall outside the magnified polygon. The data-driven factor, $\lambda_{\text{data}}$, serves as an essential geometric correction, ensuring that the final visualized fence is a faithful representation of the statistical test's outcome.
\end{remark}

\subsection{A Direct and Stable Visualization}
\label{subsec:visualization}

To address the visual instability and misinterpretation inherent in the convex hull representation (see Section~\ref{limitation:loop}), 
we replace the ``bolster'' with explicit whiskers and a direct fence visualization. This design abandons the volatile loop to faithfully map the data's tail structure, preserving details of skewness and correlation that a polygon often obscures. Our approach is defined by two principles.


\medskip
\noindent\textbf{1. Direct Fence Visualization.} The primary change is to draw the fence explicitly. This fence is the exact contour obtained by magnifying the central bag by the data-adaptive inflation factor $\lambda$ in \eqref{lambda}. This modification ensures a direct, one-to-one correspondence between the statistical rule and the graphical boundary: a point is flagged as an outlier if and only if it lies outside this line.

\medskip
\noindent\textbf{2. A Refined Whisker Implementation for Visual Clarity.} To better align with the spirit of Tukey's original box-and-whisker plot and enhance visual interpretation, we propose a refined implementation of bivariate whiskers that integrates their design and graphical composition. Our approach draws individual lines from the bag's boundary to \emph{each} outer point (point that is inside the fence but outside of the bag) in the surrounding region. 

Crucially, these whiskers are rendered with two key graphical properties. First, they are placed on a background layer, ensuring that all data points are plotted on top and remain fully visible. Second, each whisker features a transparent gradient along its length, being most transparent at its origin on the bag's boundary and becoming progressively more solid as it approaches the data point. This unique gradient serves a dual purpose: it prevents visual clutter near the dense central bag while simultaneously drawing the eye outward, visually emphasizing the most extreme observations that define the data's spread. Together, these design choices create a clear visual hierarchy: the solid bag represents the data's robust core, the gradient whiskers illustrate the extent of the non-outlying data, and the distinct fence marks the formal decision boundary.

\section{A Step-by-Step Toy Example}
\label{subsec:toy_example}

\begin{figure}[!htbp]
	\centering{
		\includegraphics[width=0.6\textwidth]{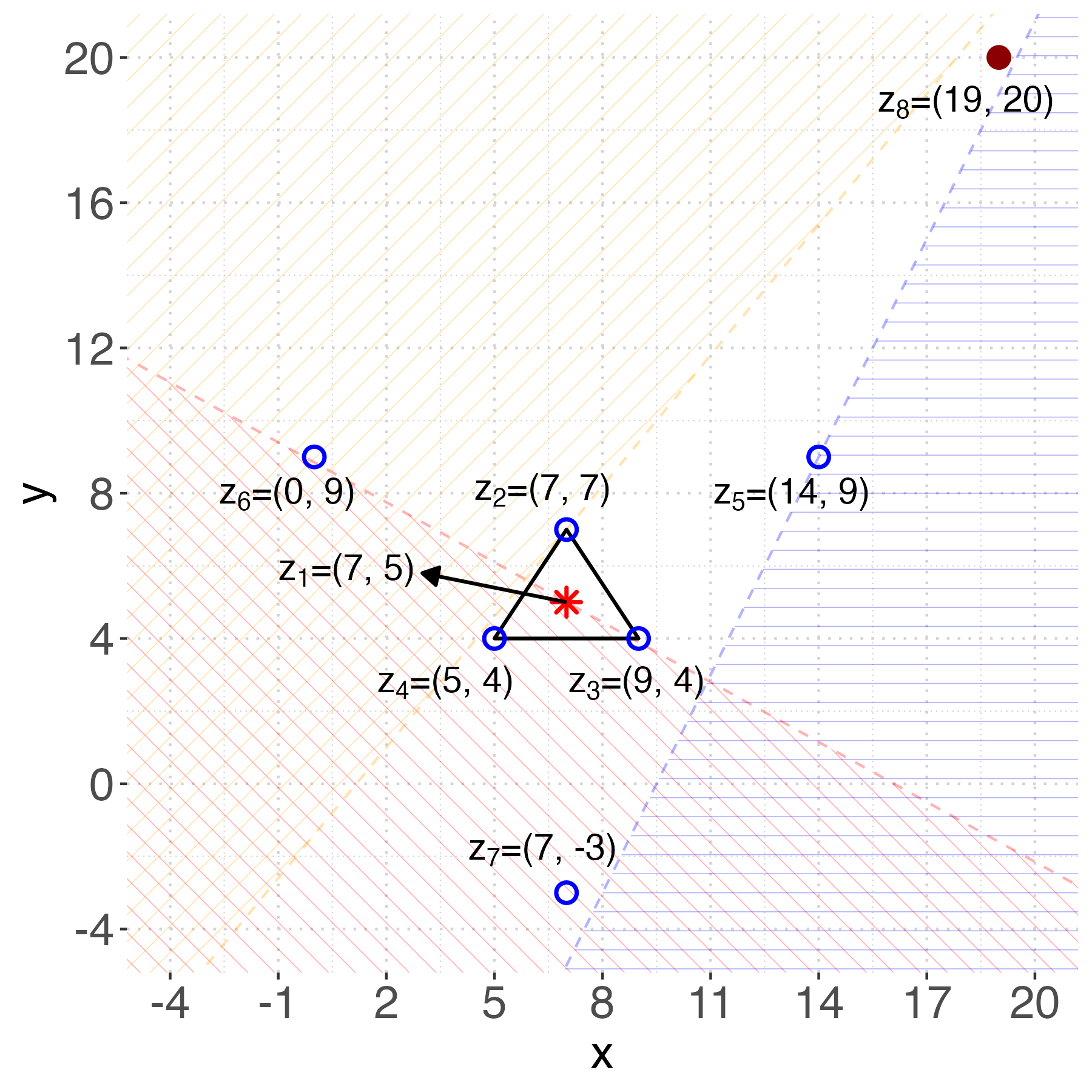}
	}
  \caption{Scatter plot and halfspace depth illustration for the toy example. The star-shaped (red) point at $\boldsymbol{z}_{1} = (7, 5)$ marks the depth median. $\boldsymbol{z}_{2}$ to  $\boldsymbol{z}_{4}$ indicate points with larger depth (the bag's core),  $\boldsymbol{z}_{5}$ to $\boldsymbol{z}_{7}$ are outer points, and the point $\boldsymbol{z}_{8}=(19, 20)$ is the extreme outlier located farthest from the majority of the data. Shaded halfspaces in the plot show example halfplanes used to compute the halfspace depth of $\boldsymbol{z}_1$, $\boldsymbol{z}_2$ and $\boldsymbol{z}_{5}$.}
	\label{fig:toy_example_hdepth}
\end{figure}

To illustrate the computation of the new bagplot, we present a simple toy example. Consider a small dataset of $n = 8$ observations as follows:
\begin{align*}
	\boldsymbol{Z} = \{&\boldsymbol{z}_i, i = 1, \ldots, 8 \} \\
	= \{& (7,5), (7,7), (9,4), (5,4), (14,9), (0,9), (7,-3), (19, 20)\}.
\end{align*}
The scatter plot of these points is shown in Figure~\ref{fig:toy_example_hdepth}. In this dataset, the observation $\boldsymbol{z}_{8}$ is well-separated from the bulk of the data.
We will walk through the pipeline to see how the procedure identifies outliers and constructs the bagplot.

\textbf{Step 1: Robust Parameter Estimation.} 
First, we estimate the location using the depth median. For each point $\boldsymbol{z}_i$, we compute its halfspace depth.
Consider the point $\boldsymbol{z}_{5}$ as an illustrative example. 
The halfplane defined by $y \leq 2x - 19$ 
 contains only the single point $\boldsymbol{z}_{5}$. Since this count of 1 is minimal across all possible halfplanes through $\boldsymbol{z}_{5}$, its halfspace depth is $1$. 
For the point $\boldsymbol{z}_2$, any closed halfplane passing through it contains at least two points, e.g., the half plane defined by $ y\geq 7(x-1)/6$ contains $\boldsymbol{z}_2$ and $\boldsymbol{z}_{6}$. 
Thus, its depth is $2$.  Similarly, we can conclude that $\boldsymbol{z}_6$, $\boldsymbol{z}_7$ and $\boldsymbol{z}_{8}$ have the depth of $1$, and $\boldsymbol{z}_3$ and $\boldsymbol{z}_{4}$ have the depth of $2$. 
The depth median is $\boldsymbol{z}_1$, which attains the maximum depth of $3$. To see this, the halfplane defined by $y\leq (-5x+80)/9$ passes through $\boldsymbol{z}_1$ but excludes $\boldsymbol{z}_3$ and $\boldsymbol{z}_6$.
The MCD estimator yields the covariance matrix $\begin{pmatrix}53/3 & 0\\ 0 & 17\end{pmatrix}$.

\textbf{Step 2: Bag Construction.} 
Next, we construct the central bag. In this toy example, we identify the set of the four deepest observations: $\boldsymbol{z}_1, \boldsymbol{z}_2, \boldsymbol{z}_3$ and $\boldsymbol{z}_4$. The bag is defined as the convex hull of these points, which forms the triangle depicted in Figure~\ref{fig:toy_example_hdepth}.

\textbf{Step 3: $p$-value Calculation.} 
Using the robust parameter estimates from step 1, we calculate the squared Mahalanobis distance, $d_i^2$, for each of the 8 data points via \eqref{eq:mahalanobis}. We then convert these distances into upper-tail probabilities using the $\chi^2_2$ reference distribution, as defined in \eqref{eq:pvalue}, to obtain a $p$-value for each point. The resulting squared Mahalanobis distances and their corresponding $p$-values are:
\begin{align*}
	\boldsymbol{d}^2 &= \{0.00, 0.24, 0.29, 0.29, 3.71, 3.71, 3.76, 21.39\}, \\
	\boldsymbol{p}\text{-value} &= \{1.00, 0.89, 0.87, 0.87, 0.16, 0.16, 0.15, 2.27\times 10^{-5}\}.
\end{align*}
The smallest $p$-value corresponds to the point $\boldsymbol{z}_{8}$, providing strong evidence that it does not follow the central data distribution.

\medskip
\noindent\textbf{Step 4: Multiple Testing Adjustment.}
With the set of $p$-values, we apply different multiple testing procedures to identify which points should be flagged as outliers. We consider three common error-control strategies:

\begin{itemize}
	\item \textbf{FWER Control:} \textcolor{black}{Using the Holm-Bonferroni method \citep{holm1979simple} at a level of {\color{black}$q=0.1$}, the procedure rejects any hypothesis corresponding to a $p$-value less than or equal to the single significant result, yielding a threshold of $t_{\text{adj}} \approx 2.27\times 10^{-5}$.}
	
	\item \textbf{FDR Control:} \textcolor{black}{Applying the BH procedure \citep{benjamini1995controlling} at {\color{black}$q=0.1$} also results in rejecting only the hypothesis for the most extreme point, giving an identical threshold of $t_{\text{adj}} \approx 2.27 \times 10^{-5}$.}
	
	\item \textbf{PFER Control:} {To control the Per-Family Error Rate at a target of {\color{black}$q=0.5$} expected false positives, we reject any hypothesis where the $p$-value is less than or equal to $q/n$. This gives a threshold of {\color{black}$t_{\text{adj}} = 0.5/8 = 0.0625$.}}
\end{itemize}

These significance thresholds are then converted back to the scale of squared Mahalanobis distance using \eqref{eq:d_adj}. This results in a critical value of $d^2_{\text{adj}} \approx 21.39$ for both FWER and FDR control, and a more liberal critical value of \textcolor{black}{$d^2_{\text{adj}} \approx 5.55$} for PFER control. In this example, the rejection set is identical across all three procedures, consisting solely of the extreme observation $\boldsymbol{z}_{8}$.

\medskip
\noindent\textbf{Step 5: Data-Adaptive Inflation Factor.}
 We now derive the inflation factor $\lambda$ using our two-stage approach. 
First, we calculate  $\lambda_{\text{stat}}$, based on the critical distances from step 4. The median of the squared Mahalanobis distances is $(0.29+3.71)/2=2$.
\begin{itemize}
	\item For FWER and FDR control, $\lambda_{\text{stat}} = \sqrt{21.39 / 2} \approx 3.27$.
	\item For PFER control, {\color{black}$\lambda_{\text{stat}} = \sqrt{5.55 / 2} \approx 1.67$.}
\end{itemize}
Next, we determine the \emph{data-driven} inflation factor, $\lambda_{\text{data}}$. Given that all three procedures rejected only the point $\boldsymbol{z}_{8}$ (the extreme outlier), the fence must be sufficiently large to geometrically enclose all remaining observations. The most extreme non-rejected point is $z_{7}$; thus, we calculate that a minimum inflation factor of $\lambda_{\text{data}} = 8$ is required to position this point on the boundary of the magnified bag.
This calculation is derived from the ratio of the distance between the depth median and $\boldsymbol{z}_{7}$ to the distance between the median and the bag's boundary along the ray connecting them. Specifically, the distance to $\boldsymbol{z}_{7}$ is $|5 - (-3)| = 8$, while the distance to the boundary intersection point $(7, 4)$ is $|5 - 4| = 1$. Finally, to satisfy both statistical and geometric criteria, we set $\lambda = \max(\lambda_{\text{stat}}, \lambda_{\text{data}})$.

\begin{itemize}
	\item For FWER and FDR, $\lambda = \max(3.27, 8) = 8$. 
	\item For PFER, $\lambda = \max(1.67, 8) = 8$. 
\end{itemize}

\begin{figure}
	\centering{
		\includegraphics[
		width=0.8\textwidth
		]{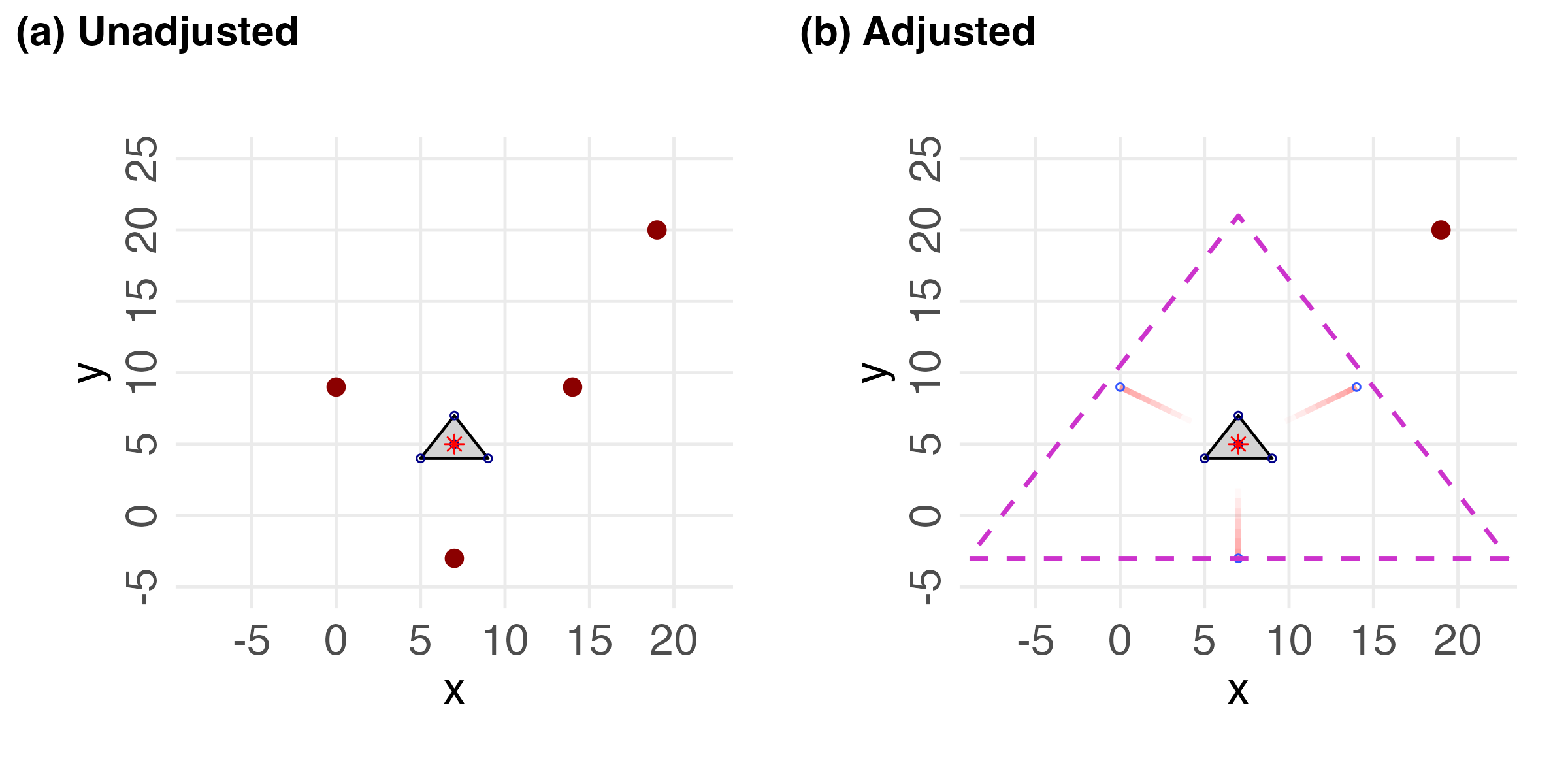}
	}
	\caption{On the Left panel is the original bagplot from \cite{rousseeuw1999bagplot}. On the right panel is the new bag-and-whisker plot. Since the inflation factor derived from the three error rates are the same, the resulting bag-and-whisker plots are identical.}
	\label{fig:toy_example_bagplot}
\end{figure}

The resulting adaptive bagplots are constructed using these final inflation factors and are shown in Figure~\ref{fig:toy_example_bagplot}.

 As intended, under all three procedures the point $\boldsymbol{z}_{8}$ is flagged as an outlier. This result aligns with our visual intuition and showcases the method's ability to produce a principled and coherent graphical summary. It is instructive to contrast this outcome with the original bagplot proposed by \cite{rousseeuw1999bagplot}. Applying the conventional fixed inflation factor of 3 to this dataset excludes four points ($\boldsymbol{z}_{5}, \boldsymbol{z}_{6}, \boldsymbol{z}_{7}\  \text{and } \boldsymbol{z}_{8}$) from the fence. Consequently, the set of non-outlying points is reduced strictly to the central bag itself, causing the visualized ``bolster'' to collapse onto the bag's boundary.

 The toy example above allows for an exact, manual verification of the methodology. However, for general datasets where $n$ is large, exact computation becomes computationally intensive. Our implementation therefore relies on established algorithmic approximations to ensure efficiency without sacrificing robust performance. The following remarks clarify the specific computational routines and software constraints relevant to the proposed method.

\begin{remark}
In the toy example with so few points, we were able to compute the halfspace depth by hand, but it is obvious that such a manual approach is not practical when $n$ is large. \cite{rousseeuw1996algorithm} proposed an efficient algorithm that can compute the halfspace depth of $n$ points in $O(n \log n)$ time. In our implementation, we adopt the same approximation method used in the \texttt{aplpack} package in R, which iterates over a fixed set of directions (e.g., 180 or 360) to find the minimum number of points in the halfspaces on either side.
\end{remark}

\begin{remark}
	In finite datasets, the point achieving the maximum halfspace depth is not always unique. Consistent with the original definition provided by \cite{rousseeuw1999bagplot}, when the set of deepest points contains more than one observation, we define the depth median as the center of gravity of this subset. Computationally, this is obtained by calculating the arithmetic mean of the coordinate vectors of all observations that attain the global maximum depth.	
\end{remark}

\begin{remark}
The MCD estimator \citep{rousseeuw1984least} is defined by the mean and covariance of the subset of $h$ observations (typically $h \approx n/2$) whose covariance matrix possesses the smallest possible determinant. To ensure statistical consistency under the multivariate normal model, this raw covariance estimate is multiplied by a specific scaling factor. Since the exact computation of the MCD is computationally intensive even for relative small $n$ as it requires the evaluation of $\binom{n}{h}$ subsets. \cite{rousseeuw1999fast} proposed the FAST-MCD algorithm to efficiently approximate the solution. In our toy example and subsequent implementation, we utilize this algorithm, including the necessary consistency scaling, via the \texttt{cov.mcd} function available in the \texttt{MASS} package in R.
\end{remark}

\begin{remark}
In this toy example, the subset of observations with a halfspace depth $\ge 2$ contains exactly $n/2 = 4$ points, allowing the bag to be defined simply as the convex hull of this set. In general practice, however, due to the discrete nature of halfspace depth, there is rarely a depth contour that strictly encloses exactly 50\% of the data. Consequently, constructing the bag requires a more sophisticated geometric interpolation strategy to approximate the 50\% central region. To handle these non-trivial cases robustly, our implementation relies on the internal routines of the \texttt{aplpack} package in R to compute the bag.
\end{remark}

\section{Simulated Data Analysis}

\label{sec:examples}

In this section, we use simulated data to demonstrate the performance of the proposed bag-and-whisker plot under controlled conditions where the ground truth is known. To provide a clear benchmark, we compare our method against the widely-used implementation of the original bagplot in the R package \texttt{aplpack}, which represents the conventional approach of a fixed inflation factor and a convex hull visualization. 
\textcolor{black}{
	We showcase three variants of our adaptive bagplot, corresponding to the control of FWER, FDR and PFER, respectively. For FWER and FDR we set the nominal levels to 0.1. For PFER we set the nominal level to 0.5 which corresponds to Chauvenet's criterion  \citep{chauvenet1863manual,lin2025tukey}.
}
Each of the following three examples is designed to test a specific aspect of the method: its statistical adaptivity to data contamination, its robustness to correlation structure, and its visual stability in the presence of heavy tails.





\begin{figure}[!htbp] 
	\centering{
		\includegraphics[width=1\textwidth,height=0.6\textheight]{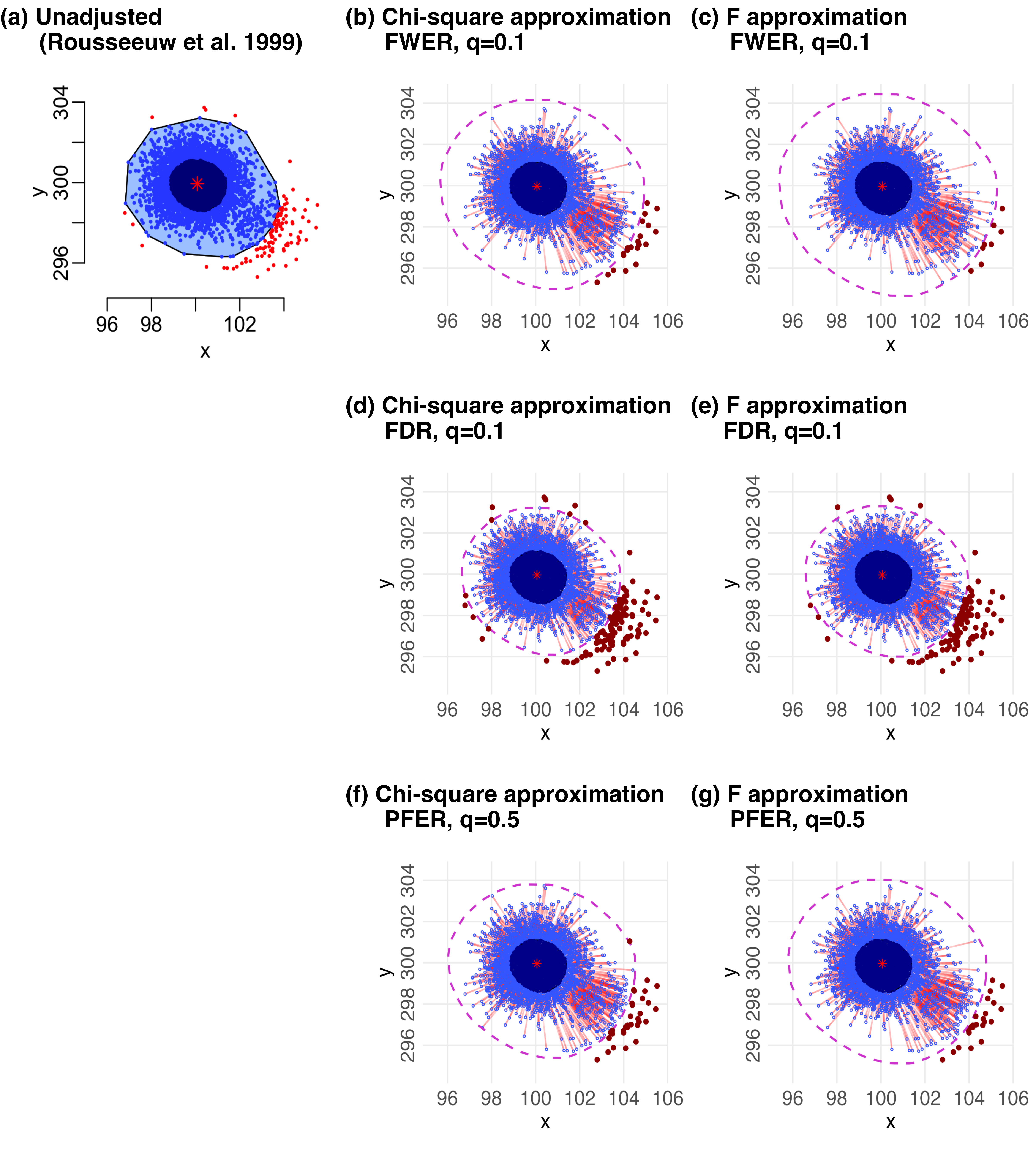}
	}
\caption{{\color{black}Bagplots for a bivariate normal mixture with a weakly separated secondary component. Panel (a): original bagplot \citep{rousseeuw1999bagplot} with fixed inflation factor and convex-hull boundary. Panels (b), (d), (f): proposed bag-and-whisker plots using the $\chi^2_2$ reference with FWER, FDR, and PFER control, respectively. Panels (c), (e), (g): corresponding plots using the scaled $F$ approximation. Key features: (i) smooth elliptical fences (dashed) that reflect the statistical threshold and avoid convex-hull artifacts; (ii) gradient whiskers extending from the bag to peripheral points, indicating the direction of the contaminating cluster; (iii) adaptive fence radii across error-control procedures, reflecting varying tolerance for false discoveries. The $F$-based fences are slightly more conservative than their $\chi^2$ counterparts. }}
	\label{fig:simu-normal_mixture}
\end{figure}

\medskip
\noindent\textbf{Example 1: Statistical Adaptivity in Independent Normal Mixture.}
Our first example tests the methods on a subtle normal mixture, designed to assess how the fences adapt to a small, contaminated component. We generate $n=5000$ points from the following model:
\begin{equation*}
	\theta_i \stackrel{iid}{\sim} \text{Bernoulli}(0.05), \quad \vec{X}_i|\theta_i \stackrel{ind}{\sim} (1-\theta_i)\mathcal{N}(\vec{\mu}_0, \mathbf{I}_2) + \theta_i \mathcal{N}(\vec{\mu}_1, \mathbf{I}_2),
\end{equation*}
where $\vec{\mu}_0 = (100, 300)^\top$, $\vec{\mu}_1 = (103, 298)^\top$ and $\mathbf{I}_2$ is the $2\times2$ identity matrix. As shown in Figure~\ref{fig:simu-normal_mixture}, the \texttt{aplpack} implementation (panel (a)) produces a boundary based on a fixed rule and a convex hull visualization that can be sensitive to the specific sample. In contrast, our bagplots (panels (b)-(g)) demonstrate clear advantages. {\color{black}The fences are smooth and elliptical, reflecting the underlying normality of the components. The different error control levels provide a nuanced view: the FWER fence is conservative, flagging only the most extreme points of the minor cluster, while the FDR fence adapts to the evidence of contamination to create tighter boundaries.  The PFER fence at level $0.5$ produces an intermediate threshold. The refined whisker construction extends toward the minor component, indicating the direction of separation while avoiding the geometric constraints of a convex hull. The fences derived from the scaled $F$ approximation are marginally more conservative than those based on the $\chi^2_2$ distribution.}

\begin{figure}[!htbp] 
	\centering{
		\includegraphics[width=1\textwidth,height=0.6\textheight]{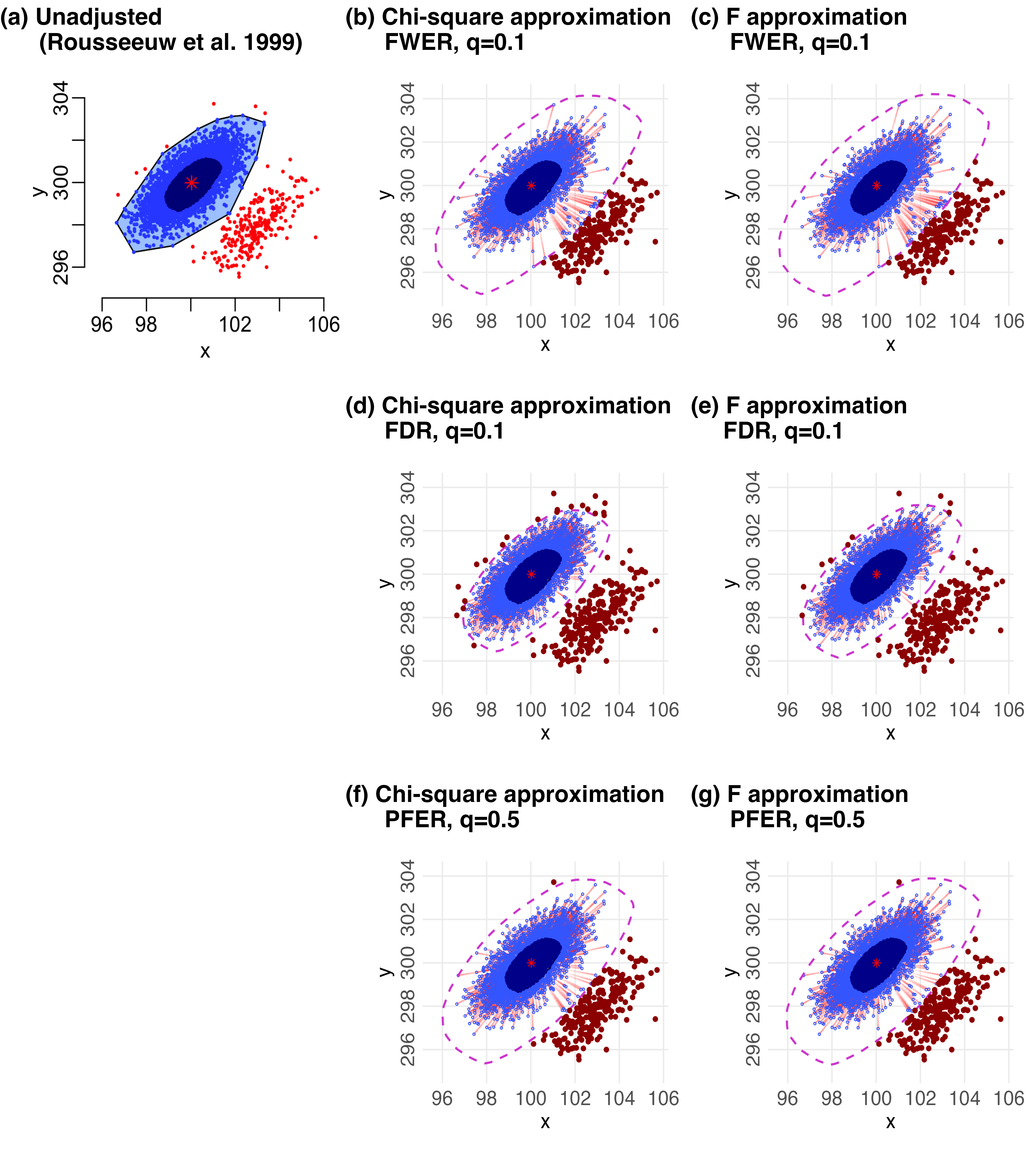}
	}
  \caption{{\color{black}Bagplots for a correlated bivariate normal mixture ($\rho = 0.3$) with a minor contaminating component. Panel (a) displays the original bagplot, where the convex hull boundary is jagged and misaligned with the underlying elliptical geometry. Panels (b), (d), (f) show the proposed method using the $\chi^2_2$ reference distribution, while panels (c), (e), (g) use the scaled $F$ reference. Note the orientation of the fences relative to the data cloud. Unlike the fixed rule in panel (a), the proposed fences are smooth and elongated, accurately reflecting the covariance structure estimated by the MCD. This alignment ensures that outlier detection respects the data's correlation. The whiskers further clarify the direction of the minor component along the major axis of correlation. }}
	\label{fig:simu-corr_normal_mixture}
\end{figure}

\medskip
\noindent\textbf{Example 2: Robustness to Correlation Structure.}
The second example employs a correlated normal mixture to demonstrate how the framework adapts to covariance structure. Data are generated from the model in the first example, but with a covariance matrix $\begin{pmatrix}1 & 0.3\\ 0.3 & 1\end{pmatrix}$. 
The  \texttt{aplpack} visualization (Figure~\ref{fig:simu-corr_normal_mixture}(a)) produces a jagged convex hull boundary which provides a noisy and unstable representation of the data's underlying correlation structure.

The bag-and-whisker plot excels in this scenario. Because our pipeline uses the robust MCD estimator for the covariance matrix, both the central bag and the resulting fences naturally elongate along the main axis of correlation (Figure~\ref{fig:simu-corr_normal_mixture}(b)-(g)). This yields an interpretable elliptical-like contour that correctly reflects the data's covariance structure, providing a much more accurate and stable boundary for outlier detection in correlated multivariate settings.

\begin{figure}[!htbp] 
	\centering{
		\includegraphics[width=1\textwidth,height=0.6\textheight]{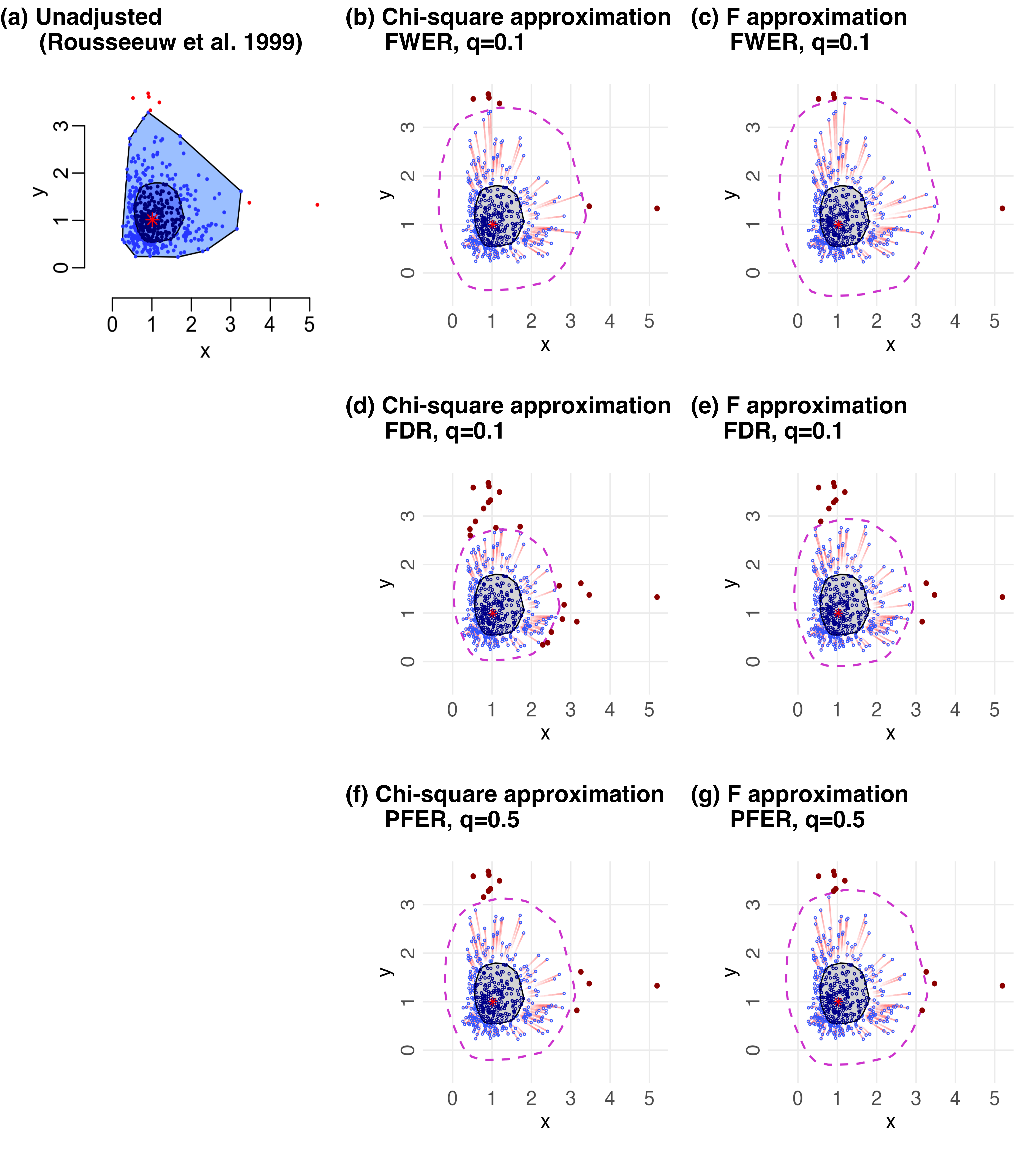}
	}
  \caption{{\color{black}Bagplots for simulated heavy-tailed log-normal data without true contamination, illustrating visual stability in the presence of skewness and heavy tails. Panel (a) shows the original bagplot, where the convex hull produces an irregular, jagged boundary driven by peripheral points. Panels (b), (d), (f) ($\chi^2_2$ reference) and Panels (c), (e), (g) (scaled $F$ reference) yield smooth and stable fences that faithfully reflect the underlying distribution. The gradient whiskers clarify directional spread without visual clutter. }}
  
	\label{fig:simu-lognorm}
\end{figure}

\medskip
\noindent\textbf{Example 3: Visual Stability with Heavy-Tailed Data.}
Our final example highlights the importance of visual stability. We generate 500 points with independent $x$ and $y$ coordinates drawn from log-normal(0,0.5) distribution, which is skewed and heavy-tailed but has no true contaminant outliers. Figure~\ref{fig:simu-lognorm}(a) demonstrates the primary drawback of using a convex hull for visualization in such cases. The \texttt{aplpack} boundary is highly irregular and jagged, its shape dictated by the handful of most extreme observations. This creates a visually unstable representation that does not reflect the smooth underlying data distribution.

In contrast, the fences of the bag-and-whisker plot (Figure~\ref{fig:simu-lognorm}(b)-(g)) remain smooth and stable. By constructing the fence directly from a magnified version of the robustly estimated bag, our method is insensitive to small perturbations of peripheral points.

\section{Real-World Data Applications}
\label{sec:real-data}
\begin{figure}[!htbp] 
\centering{
\includegraphics[width=1\textwidth,height=0.6\textheight]{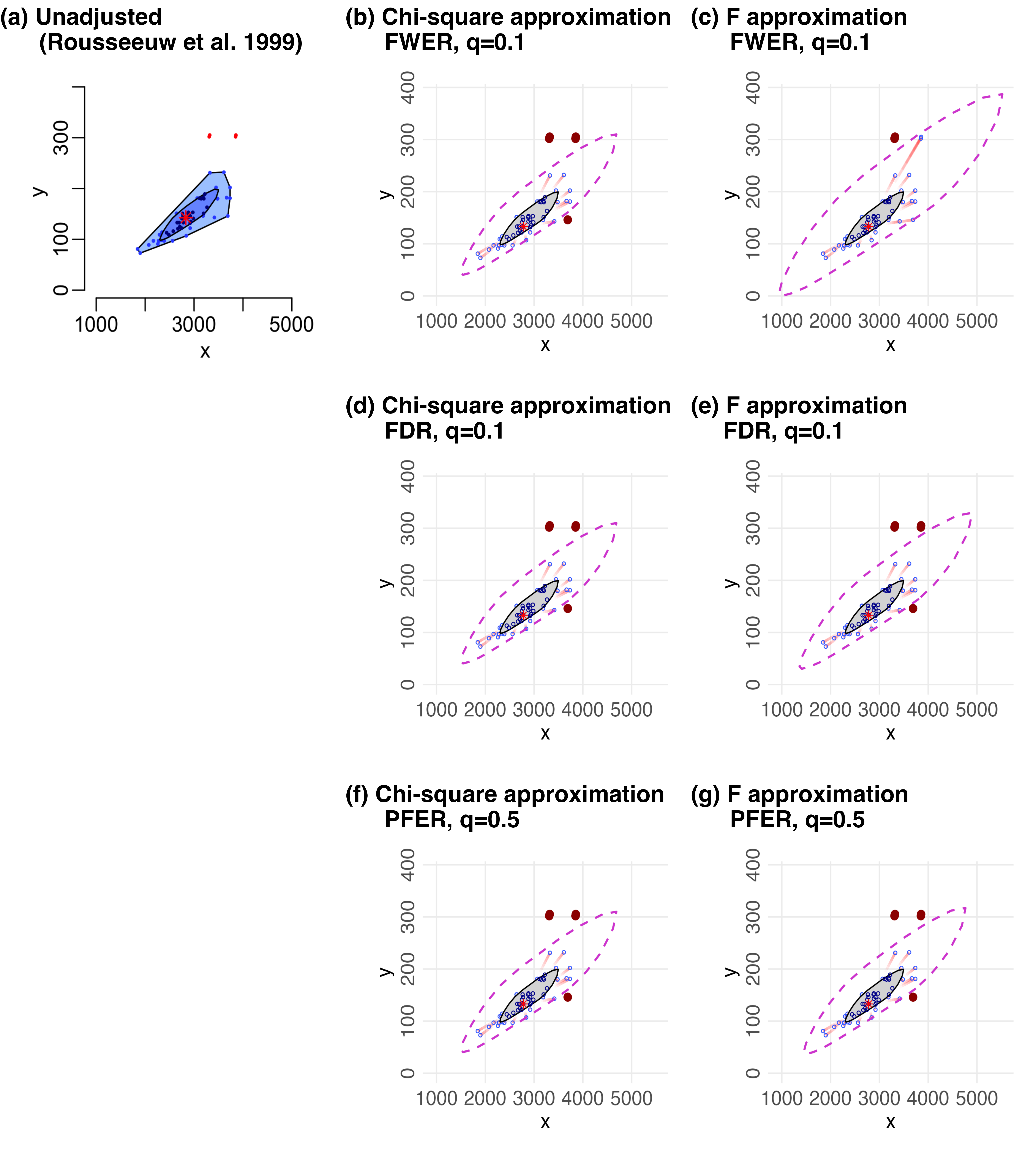}
}
\caption{{\color{black}Reanalysis of the data from Figure 1 of  \citet{rousseeuw1999bagplot}. Panel (a) displays the original implementation, where the outer boundary is constructed as the convex hull of non-outlying points. This yields a jagged, geometrically unstable loop that can contract in sparse regions. Panels (b)--(g) present the proposed method, which replaces the hull with a directly rendered fence obtained by magnifying the central bag, supplemented by gradient whiskers.}}
\label{fig:rd-fig1_rousseeuw1999bagplot}
\end{figure}

\begin{figure}[!htbp] 
\centering{
\includegraphics[width=1\textwidth,height=0.6\textheight]{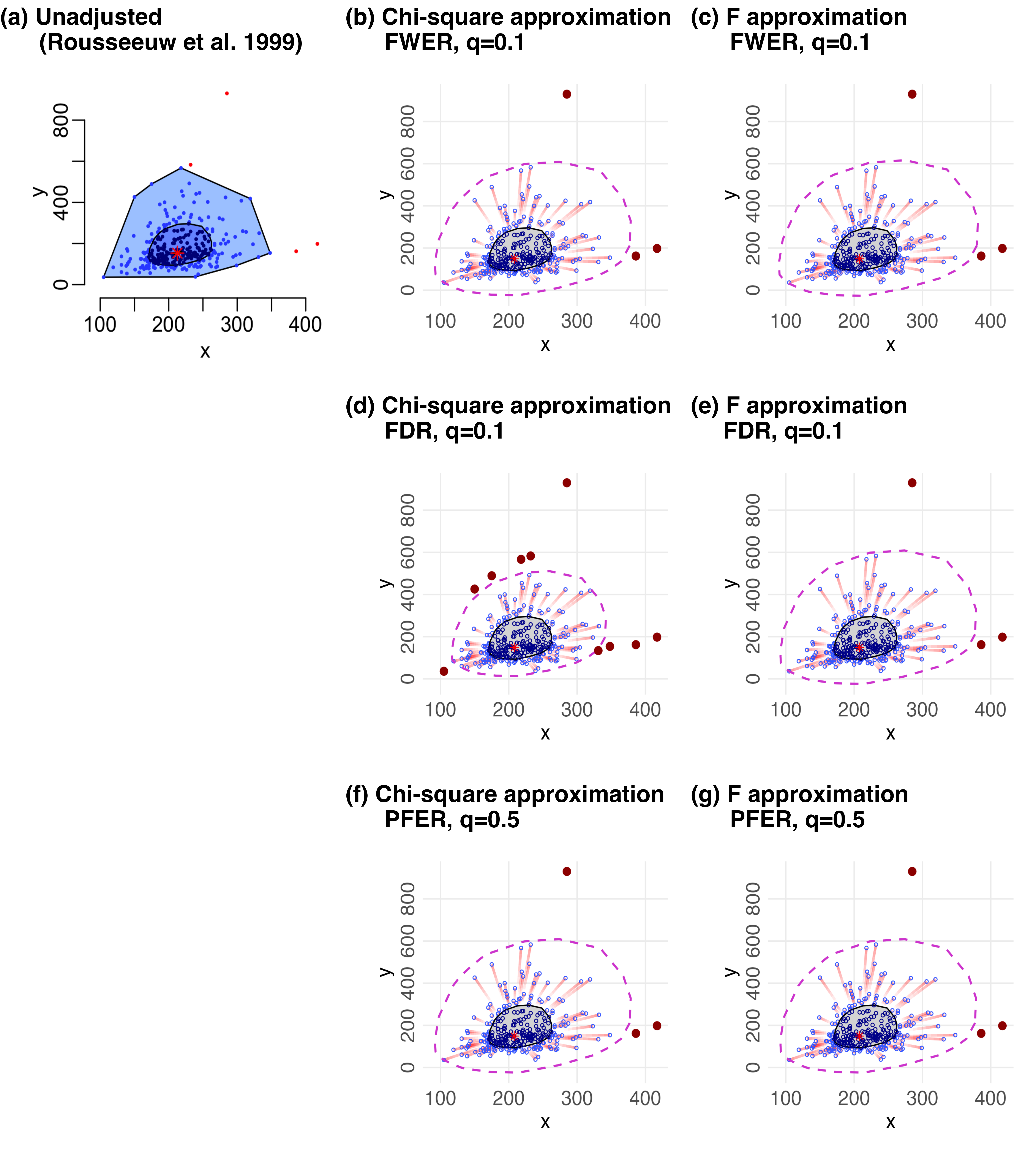}
}
\caption{{\color{black}Reanalysis of the data from Figure~3(a) of \citet{rousseeuw1999bagplot}. Panel (a) displays the original convex-hull loop, yielding a jagged, vertex-driven boundary. Panels (b)--(g) replace the hull with a directly rendered fence. This eliminates polygonal artifacts, producing smooth and stable fences that more accurately reflect the data's structure.}}
\label{fig:rd-fig3a_rousseeuw1999bagplot}
\end{figure}

\begin{figure}[!htbp] 
\centering{
\includegraphics[width=1\textwidth,height=0.6\textheight]{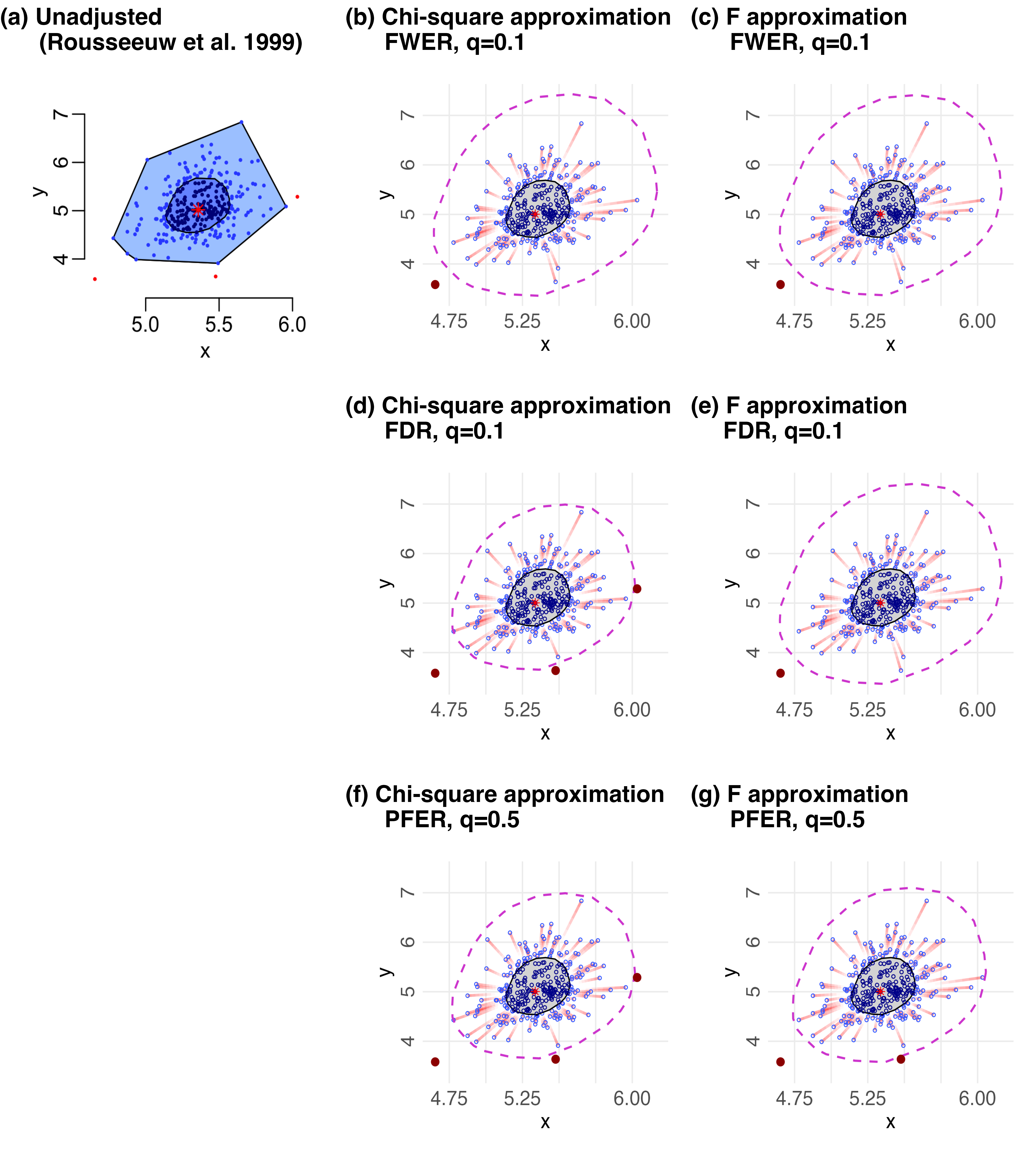}
}
\caption{{\color{black}Reanalysis of the data from Figure~3(b) of \citet{rousseeuw1999bagplot}. Different choices of error metrics yield different inflation factors and outliers. The use of scaled $F$-distribution results in more conservative fences.}
\label{fig:rd-fig3b_rousseeuw1999bagplot}}
\end{figure}

\begin{figure}[!htbp] 
\centering{
\includegraphics[width=1\textwidth,height=0.6\textheight]{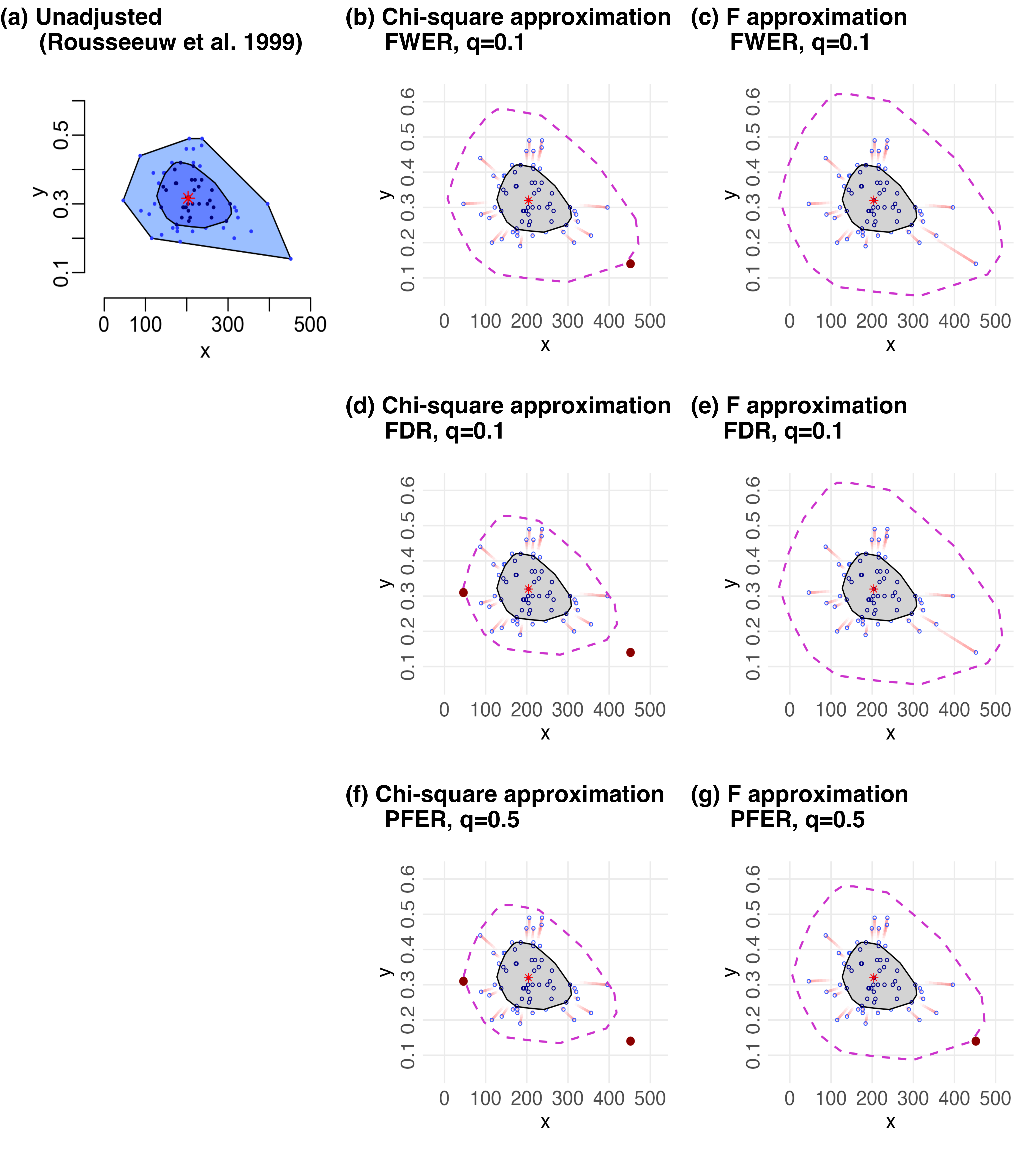}
}
\caption{{\color{black}Reanalysis of the data from Figure~6 of \citet{rousseeuw1999bagplot}. Similar as the previous figures, different choices of error metrics yield different inflation factors and outliers. The use of scaled $F$-distribution results in more conservative fences again. }}
\label{fig:rd-fig6_rousseeuw1999bagplot}
\end{figure}

This section demonstrates the practical benefits of the new bagplot using real data. For a fair comparison, we revisit several of the classic datasets originally used to introduce the bagplot in \citet{rousseeuw1999bagplot}. This allows for a direct comparison of our method's output with the visualizations produced by the conventional approach, as implemented in the R package \texttt{aplpack}.
Specifically, we re-examine the data from Figure~1, Figure~3(a), Figure~3(b) and Figure~6 of the original paper in our Figures~\ref{fig:rd-fig1_rousseeuw1999bagplot}, \ref{fig:rd-fig3a_rousseeuw1999bagplot}, \ref{fig:rd-fig3b_rousseeuw1999bagplot} and \ref{fig:rd-fig6_rousseeuw1999bagplot}, respectively.

\medskip
\noindent\textbf{Enhancing Visual Stability and Interpretability.}
We revisit the automobile dataset from Figure~1 of \citet{rousseeuw1999bagplot}, {\color{black}which records vehicle weight ($x$, in pounds) and engine displacement ($y$, in cubic inches) for $n=60$ cars \citep{chambers1993statistical}. The data exhibit moderate positive correlation with a distinct upper-right cluster of high-displacement vehicles. The analytic objective of this reanalysis is to assess boundary stability in a moderate-sample setting where peripheral points typically dominate convex-hull geometry.} Figure~\ref{fig:rd-fig1_rousseeuw1999bagplot} contrasts the original \texttt{aplpack} implementation with the proposed method. Panel~(a) displays the conventional convex-hull loop, which contracts inward in the sparse upper-right region, producing an artifact that misrepresents the underlying depth geometry. Panels~(b)--(g) replace this loop with a smooth, convex fence derived from the data-adaptive inflation factor. The gradient whiskers extend toward the high-displacement cluster, clarifying the direction of the tail without obscuring the central bag. This revision eliminates the visual instability inherent in the bolster representation while preserving the original identification of four extreme vehicles as outliers.

We next examine the plasma concentration data from Figure~3(a) of \citet{rousseeuw1999bagplot}, {\color{black} which comprises measurements of cholesterol ($x$, mg/dl) and triglycerides ($y$, mg/dl) for $n=320$ patients with arterial narrowing \citep{hand1994handbook}. The distribution is right-skewed with a heavy upper tail, a structure that typically challenges polygonal boundary definitions.} Figure~\ref{fig:rd-fig3a_rousseeuw1999bagplot} presents the results. While both our method and the original bagplot identify the same set of clear outliers, our directly-rendered fence offers a more nuanced boundary. It smoothly delineates the space between the majority and outliers, enhancing the visual representation of the data's spread, whereas the convex hull in panel~(a) is again jagged and sensitive to the specific locations of a few points.


\medskip
\noindent\textbf{Providing Statistical Flexibility in Outlier Detection.}
Beyond visual clarity, our framework provides a more powerful and flexible statistical tool. The classic bagplot offers a single, static assessment based on a fixed rule. 
To illustrate the benefit of our adaptive approach, we re-analyze two datasets from the original paper. {\color{black}Figure~\ref{fig:rd-fig3b_rousseeuw1999bagplot} revisits the log-transformed plasma cholesterol and triglyceride concentrations for $n=320$ patients with evidence of narrowing arteries \citep{hand1994handbook}. The log-transformation was originally applied to address the pronounced right-skewness in these chemical concentration measurements. Figure~\ref{fig:rd-fig6_rousseeuw1999bagplot} examines the polychlorinated biphenyl (PCB) concentration and eggshell thickness for $n=65$ Anacapa pelican eggs \citep{hand1994handbook}, a dataset originally used to study the environmental impact of industrial pollutants on bird reproduction.}

In both examples, our bag-and-whisker plot identifies a varying number of outliers depending on the chosen error metric, exhibiting far greater flexibility than the original bagplot. This variability is not an inconsistency but a key feature of our framework. It empowers the analyst to choose an error metric and level that matches their scientific goals. The proposed  bag-and-whisker plot effectively and transparently visualizes the results of this tailored statistical decision, offering a more nuanced and powerful tool for data exploration.

\section{Conclusion}
\label{sec:conclusion}
In this paper, we introduced a new bivariate boxplot, namely the  bag-and-whisker plot, to address the long-standing statistical and graphical limitations of the original bagplot. 
Our contribution is twofold. First, we developed a general $p$-value pipeline that recasts outlier detection as a multiple testing problem, generating a data-adaptive fence whose stringency is aligned with formal error control principles. Second, and perhaps more significantly for data visualization, we replaced the unstable convex hull (the bolster) with a refined visualization that renders the fence directly and employs gradient whiskers. This ensures a one-to-one correspondence between the method's statistical rule and its graphical output. The result is not merely an adaptive update, but a structural evolution of the method, offering a principled and interpretable standard for modern exploratory data analysis.


We emphasize that the bag-and-whisker plot should be understood as a tool for exploratory data analysis rather than formal inference. While its fences are constructed in alignment with multiple testing principles, the procedure does not rigorously control error rates due to its reliance on estimated null parameters. Its main contribution is conceptual, providing a coherent framework for developing and evaluating principled outlier detection rules for graphical exploration, rather than serving as a formal hypothesis test itself.

Acknowledging this exploratory role, the framework nevertheless opens several promising avenues for future research. While this paper focused only on the bivariate case for clarity of visualization, the underlying statistical methodology is inherently general and extends directly to higher dimensions. The Mahalanobis distance, robust estimators, and the multiple testing pipeline are well-defined for any dimensions. The primary challenge in higher dimensions is visualization, and future work could explore effective renderings of 3D bags or apply the framework within scatterplot matrices. Furthermore, the modularity of our pipeline invites extensions to non-normal data by substituting more appropriate reference distributions. By transforming a classic graphic into a dynamic and statistically transparent tool, the bag-and-whisker plot is better suited for the rigor and complexity of modern data exploration.

\section*{Data Availability Statement}
{\color{black}The source code used to generate the numerical results and figures presented in this paper is publicly available in the GitHub repository at \url{https://github.com/seanq31/BagWhiskerPlot}. The R package \texttt{BagWhiskerPlot} that implements the new bag-and-whisker plots is available on CRAN.}

\bibliographystyle{apalike}
\bibliography{bibliography.bib}

\end{document}